\documentclass[11pt]{article}
\usepackage{amsmath}
\usepackage{amsthm}
\usepackage{amsfonts}
\usepackage{bbm}
\usepackage{graphicx}
\usepackage{sectsty}
\RequirePackage{setspace}
\usepackage{appendix}
\usepackage{float}
\usepackage{url}
\usepackage{caption}
\usepackage{lineno}
\usepackage[round]{natbib}
\usepackage[table,xcdraw]{xcolor}
\usepackage{float}

\newcommand{\compl}{{\mathbb C}}

\newcommand{\captionfonts}{\footnotesize}
\makeatletter 
\long\def\@makecaption#1#2{%
\vskip\abovecaptionskip
\sbox\@tempboxa{{\captionfonts #1: #2}}%
\ifdim \wd\@tempboxa >\hsize
{\captionfonts #1: #2\par}
\else
\hbox to\hsize{\hfil\box\@tempboxa\hfil}%
\fi
\vskip\belowcaptionskip}
\makeatother 

\begin{document}

\title{\bf Quantum Measurement, Entanglement  and the Warping Mechanism of Human Perception}

\author{Diederik Aerts$^*$, Jonito Aerts Argu\"elles\footnote{Center Leo Apostel for Interdisciplinary Studies, 
        Vrije Universiteit Brussel (VUB), Pleinlaan 2,
         1050 Brussels, Belgium; email addresses: diraerts@vub.be,diederik.johannes.aerts@vub.be} 
        $\,$ and Sandro Sozzo\footnote{Department of Humanities and Cultural Heritage (DIUM) and Centre CQSCS, University of Udine, Vicolo Florio 2/b, 33100 Udine, Italy; email address: sandro.sozzo@uniud.it}              }
        
\date{}

\maketitle

\begin{abstract}
\noindent 
We prove that the quantum measurement process contains the same warping mechanism that occurs in categorical perception, a phenomenon ubiquitous in human perception. This warping causes stimuli belonging to the same category to be perceived as more similar, while stimuli belonging to different categories are perceived as more different.
As a result of a detailed study of the quantum measurement using the Bloch representation, we identify the natural metric for pure states, namely the Fubini Study metric, and the natural metric for density states, namely the trace class metric. The warping mechanism of categorical perception is then manifested, when the distances between pure states, playing the role of stimuli for the quantum measurement, are warped into the distances between density states, playing the role of percepts for quantum measurement. We work out the example of a two-dimensional quantum model, a qubit, with `light' and `dark' as the two eigenstates, and show how the typical contraction and dilation warping of human perception manifest themselves in this example of the quantum measurement model of light and dark. 
\end{abstract}

\section{Introduction \label{introduction}}
In previous work we have shown, within the research field called `quantum cognition' \citep{aertsaerts1995,aertsgabora2005b,busemeyeretal2006,aerts2009a,aerts2009b,bruzagabora2009,aertssozzo2011,aertsetal2012,busemeyerbruza2012,havenkhrennikov2013,khrennikov2014,dallachiaraetal2015,pothosetal2015,blutnerbeimgraben2016,moreirawichert2016,yearsley2017,aertsarguelles2018,busemeyeretal2019,surovetal2019,aertsarguellessozzo2020,aertsbeltran2022a}, how human perception can be considered as a form of quantization \citep{aertsaertsarguelles2022}, due to the ubiquity of the warping mechanism in categorical perception \citep{harnad1987,goldstonehendrickson2010}. When we use the expression `quantization' we mean the phenomenon described by quantum mechanics in connection with the entities of physical reality, where, for example, photons are the quanta of light. 
A good and well-known example of the phenomenon of categorical perception is how humans perceive colors. Although the frequencies of visible light vary continuously from red, the lowest frequency visible to humans, to violet, the highest frequency, we perceive seven different colors, red, orange, yellow, green, blue, indigo, and violet. Our human perception divides this continuous region into seven colors, and the division is due to a systematic warping of the distances that distinguish frequencies from each other. Indeed, the warping of categorical perception causes distances belonging to frequencies leading to the same color to be contracted, while distances of frequencies leading to different colors to be dilated. This systematic warping also creates the colors in its initial stage, starting from stimuli that vary continuously and evenly over the frequency range of visible light. 

Having given this example of how the mechanism of categorical perception causes the clumping of groups of frequencies in areas we call `colors', it is important to mention that categorical perception is present in any form of human perception \citep{brunerpostman1949,libermanetal1957,libermanetal1967,lane1965,eimasetal1971,lawrence1949,berlinkay1969,daviesetal1998,davidoff2001,regierkay2009,havywaxman2016,hessetal2009,disaetal2011,sidman1994,schustermanetal2000,rosch1973,collieretal1973}. Because the phenomenon is not widely known outside the domain of psychology that studies it, we give a brief historical description of the phenomenon in Section \ref{categoricalperceptionandcolors}, mainly to show how fundamental it is in terms of human perception. 

In \citet{aertsaertsarguelles2022}, we identified the clumping as a consequence of categorical perception as a form of quantization that occurs in quantum mechanics for physical reality.  It is within a general study of the quantum cognition domain that we frame this identification, thereby expanding this field of research. Indeed, quantum cognition mainly introduces quantum probabilities to build models of human decisions that often better describe experimental data than models using classical Kolmogorovian probabilities, such as Bayesian models (see \citet{huangetal2025} for an up-to-date detailed and systematic overview of the empirical results of quantum cognition, and also the textbook \citet{busemeyerbruza2012} has been updated this year). Although two of the authors of the present article have been long-time researchers in quantum cognition \citep{aertsaerts1995,aertsgabora2005a,aertsgabora2005b,aertssozzo2011,aertssozzo2014}, we only recently became aware of the ubiquity of the categorical perception mechanism in human cognition, which makes it clear that along with quantum probabilities, another quantum phenomenon, namely quantization, is also present in human cognition. 

The research we propose in this article contains a direct identification of how the categorical perception mechanism leads to quantization. Indeed, we will show that the quantum measurement model itself, the so-called `collapse of the wave function', intrinsically contains the warping of categorical perception. Since in every model proposed from quantum cognition for some aspect of human cognition the structure of the quantum measurement is also present -- indeed it is by means of the collapse procedure that quantum probabilities are renormalized --- the warping similar to the one of categorical perception is introduced into the model. However, we will show that there is also a deep connection of the distortion mechanism of categorical perception with the phenomenon of entanglement, and more specifically with the entanglement of the measuring device with the considered quantum entity. We will show that the insights we present in this article, connecting the measurement procedure with entanglement of the measuring device during the short phase of a measurement and with the warping of categorical perception, sheds new light on the so-called quantum measurement problem, the structure and dynamics of human perception, and the phenomenon of quantization. In \citet{aertsarguelles2025}
we argue, using the analysis in this article, that we can consider colors as `quanta' of light for human vision in a similar way as we consider photons as quanta of light for the physical measurements on light by which photons are identified. 

The content of this article can be summarized as follows.
In Section \ref{categoricalperceptionandcolors} we give a brief historical account of the phenomenon of categorical perception and its ubiquity in human perception. In Section \ref{quantummeasurement} we introduce all the necessary elements regarding the measurement model of quantum mechanics and introduce the necessary analyzes to prove in Section \ref{categoricalperceptionquantummeasurement} that the typical warping present in categorical perception is intrinsically part of the structure of quantum measurement.

\section{Human Perception \label{categoricalperceptionandcolors}} 
In this section we describe the phenomenon of categorical perception in its historical context. Around 1950, events occurred that would bring the phenomenon of categorical perception to light in research on speech perception. Work was being done on a speech device, the `pattern playback machine', which should make it possible to automatically convert texts into spoken form so that, for example, blind people could read with it. To realize such a device, speech had to be analyzed very thoroughly. Alvin Liberman and his team worked on this realization and noticed that in generating a continuum of evenly distributed consonant-vowel syllables with endpoints reliably identified as `b', `d' and `g', there is a point of rapid decrease in hearing the sound as a `b' to hearing it as `d'. At a later point, there is a rapid transition from `d' to `g' \citep{libermanetal1957}. This finding was historically the first experimental identification of what would later be called `categorical perception'. Liberman came up with an original hypothesis aimed at explaining why people perceive an abrupt change between `b' and `p' in the way speech sounds are heard, as opposed to what happens with a synthetic morphing apparatus that produces the sounds with a continuous transition. His hypothesis was that this phenomenon is due to a limitation of the human speech apparatus, which, due to the muscular nature of its construction, would be unable to produce continuous transitions. Due to the way people produce these sounds with the muscles of the mouth and tongue and the larynx, they would not be able to pronounce anything between `b' and `p'. Therefore, when someone hears a sound from the synthetic vocal apparatus, they try to compare that sound with what they would have to do with their vocal apparatus to produce this sound. Since a human vocal apparatus can only produce `b' or `p', all continuous synthetic stimuli will be perceived as `b' or `p', whichever is closest. 

This hypothesis of Liberman also became the basis of what is now called the `motor theory of speech perception'. This theory assumes that people perceive spoken words by recognizing the gestures in the speech channel used to pronounce them, rather than identifying the sound patterns that produce speech \citep{libermanetal1967}. However, the theory was questioned when research showed that `identification' and `discrimination' of stimuli not associated with speech behave in a similar way \citep{lane1965}. It was also found that children, even before they could speak, exhibited the specific effect associated with speech perception \citep{eimasetal1971}. Step by step, it became clear that categorical perception was a much more general phenomenon than the effect associated with speech. The connection was also made with earlier findings about how stimuli are organized. An important role in this was played by Lawrence's experiments and his hypothesis of `acquired distinctiveness', a phenomenon that turned out to be a very fundamental effect of perception. The hypothesis of acquired distinctiveness shows that stimuli to which one has learned to respond differently become more distinctive, whereas stimuli to which one has learned to respond the same become more similar. Both effects are at work in people in a multitude of perceptions.

We already mentioned as an example of categorical perception the seven colors present as discrete structures in human visual perception relative to the continuum of stimuli of light frequencies in the visual spectrum of electromagnetic radiation. 
It is research on colors that inspired Eleanor Rosch to develop the prototype theory for human concepts \citep{rosch1973}. The theory states that there exists a central element for a concept, called the `prototype', relative to which the exemplars of the concept can be placed within a graded structure. Rosch came to her idea while studying the categorical structure for colors and basic shapes among the Dani, a people living in Papua New Guinea. They originally had only two words to denote colors, one meaning {\it Bright} and the other {\it Dark}. \footnote{As in other articles by our Brussels research group, when concepts appear as subjects in the text, we denote them by writing them with a capital letter and in italics}. The Dani also had no words in their language for basic shapes such as {\it Circle}, {\it Square} and {\it Triangle}. Rosch examined whether there was a difference in learning between two groups of Dani volunteers, one group learning colors and basic shapes starting with stimuli that are prototype colors and prototype basic shapes, while the other group learned starting with stimuli that are different distortions of these prototypes. In a significant way, it was shown that for both colors and basic shapes, learning was more qualitative for the group that learned starting with the prototype stimuli. This evaluation of `more qualitative learning' took into account the three features of how this can be measured, namely the ease of learning sets of categories when a particular type was the prototype, the ease of learning individual types within sets, and by rank of rating types as the best example of categories, when the prototypes of both colors and shapes were the stimuli in the learning process. 

After Rosch's work, mathematical prototype models based on fuzzy sets were developed for concepts and experimentally tested, and a subdiscipline in psychology emerged that studied concepts from this approach \citep{collieretal1973,rosch1975,roschetal1976,smithmedin1981,medinetal1984,geeraerts2001,johansenkruschke2005}. It is because the pet-fish problem was identified, a problem that arises when combining concepts \citep{oshersonsmith1981}, that one of the authors of this paper, along with a then Ph.D. student, came up with the introduction of quantum mechanical probabilities to describe the combination of concepts, since the pet-fish problem directly showed that fuzzy sets fall short in their modeling \citep{aertsgabora2005a,aertsgabora2005b}. Because it is important for some aspects of the problem we are addressing in the current article, we want to consider in more detail what this pet-fish problem is. The question asked was `how come {\it Guppy} is not a typical example of a {\it Pet}, nor a typical example of a {\it Fish}, but a very typical example of a {\it Pet Fish}'. When, in our Brussels quantum research group, we were faced with \citet{oshersonsmith1981}, we had already been working for several years on the introduction of quantum probabilities in human cognition \citep{aertsaerts1995}, and the nature of the problem posed was reminiscent of the existence of quantum interference between {\it Pet} and {\it Fish} when combined into {\it Pet Fish} \citep{aertsgabora2005a,aertsgabora2005b}. The detailed quantum framework in which to express this interference was not yet completely clear in these early years, and became incrementally clear the following years
\citep{aerts2009a,aerts2009b}. Their relevance to quantum information science was demonstrated by also modeling data obtained from the World Wide Web in a similar way with a quantum model \citep{aertsetal2012}.

Research in the years that followed showed that there were many more similarities between the structure and dynamics which concepts of human language are subject to and the structure and dynamics of quantum entities as described by quantum mechanics. For example, the well-known phenomenon of quantum entanglement also occurs with concepts, that is, it is possible to combine concepts such that experiments on these combinations violate Bell's inequalities, which is the experimental test for the presence of entanglement \citep{aertssozzo2011,aertssozzo2014}.

Although, having recently studied her early work with colors and basic shapes \citep{roschheider1971,roschheider1972,mervisetal1975}, we believe that the phenomenon of categorical perception influenced Rosch in proposing prototype theory for human concepts, we found no references to it in later works on prototype theory. This is probably due to the prototype researchers' lack of attention to what takes place during human perception. It is also because in our quantum mechanical modeling of human concepts we were mostly inspired by what the prototype researchers were working on, that we only recently learned about the phenomenon of categorical perception and how it is ubiquitous in human perception. Immediately we saw a connection to the phenomenon of quantization, as it occurs in quantum mechanics, which prompted the writing of \citet{aertsaertsarguelles2022}. However, in addition, it also sheds a very explanatory light on the problem of combining concepts, the pet fish problem, and why prototype theory fails to formulate a satisfactory answer to it. 
Indeed, categorical perception, as the example of colors illustrates well, forms compactions, and for concepts denoting these compactions, prototype theory works well.
If two such concepts combine, then, given knowledge about the phenomenon of categorical perception, it is obvious that generally the combination does not contain a prototype because the combination has not been the target of the densification leading to such a compaction brought about by the mechanism of categorical perception. The comparison we made with quanta becomes even clearer with it, two different quanta will interfere when they combine in interaction to become a physical entity, and similarly that happens for two different prototype containing concepts, when they combine in interaction they will interfere to form a new conceptual entity, which, however, generally does not contain a prototype.
In Section \ref{quantummeasurement}, we examine the connection between categorical perception and quantum measurement, more specifically, we look in detail at the situation of a two-dimensional quantum entity and show how its measurement dynamics is the underlying ground for the warping mechanism at work in the phenomenon of categorical perception. 

\section{Quantum Measurement \label{quantummeasurement}}
We will work with the Bloch representation of a two-dimensional quantum entity that we will analyze with the intention of identifying the categorical perception mechanism at work. Traditionally, the Bloch model is only a representation of the set of quantum states, pure states, and density states, also called mixed states, of a two-dimensional quantum entity, however, in the 1980s an extension of the Bloch model was worked out in which also the measurements could explicitly be represented \citep{aerts1986}. In later years, the original extension was elaborated and perfected \citep{aertssassolidebianchi2014,aertssassolidebianchi2016}, and since this extension also contains a specification of the change provoked by the measuring apparatus on the state of the measured upon entity, we will use it for our analysis, because it will more easily allow us to identify the categorical perception mechanism. We introduce the extended Bloch model for a two-dimensional quantum entity step by step, explaining details as they arise, the intention being to make the content of our article accessible to readers who are not experts in quantum physics. In parallel, we also introduce the elements of the traditional complex Hilbert space formalism of a two-dimensional quantum entity. That we speak of a `two-dimensional' quantum entity, by the way, refers to that its Hilbert space has dimension 2. Within the discipline of quantum computation, a two-dimensional quantum entity is called a `qubit'. 
 
A two-dimensional complex Hilbert space $\mathcal{H}$ is a vector space of dimension 2 over the complex numbers with an inner product, which is a map, conjugate linear in its first variable and  linear in its second, to the set $\compl$ of complex numbers 
\begin{eqnarray}
    &&\langle \ | \ \rangle: \mathcal{H} \times \mathcal{H} \rightarrow \compl \\ 
    &&\langle a\psi_1 + b\psi_2| \psi \rangle = a^* \langle \psi_1| \psi \rangle + b^* \langle \psi_2| \psi \rangle \\ 
    && \langle \psi| a\psi_1 + b\psi_2  \rangle = a \langle \psi| \psi_1  \rangle + b \langle \psi| \psi_2  \rangle  \\ 
    && \langle \psi_1| \psi_2 \rangle = \langle \psi_2| \psi_1 \rangle^* 
\end{eqnarray}
The inner product allows to express when two states are orthogonal to each other, namely, if their inner product is equal to zero. It also introduces a norm on the Hilbert space, defined in such a way that the inner product of a vector with itself is the square of the length of this vector.

\begin{figure}[h!]
\begin{center}
\includegraphics[width=10cm]{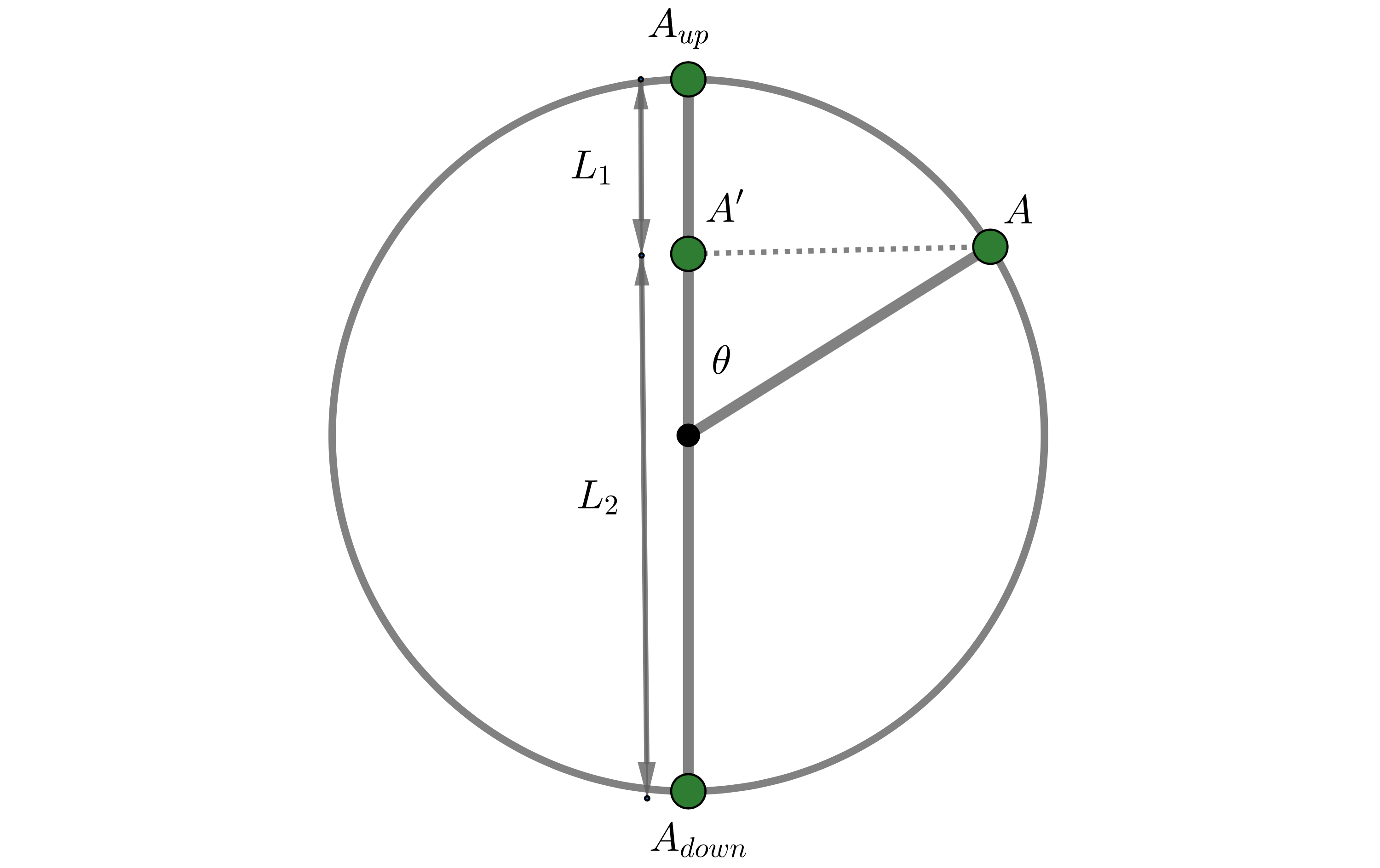}
\end{center}
\caption{A representation of our extended Bloch model for a two-dimensional quantum entity. A little ball is placed in a point $A$ on the surface of a sphere and this represents the state of the entity in $A$ before the measurement. On a central axis of the sphere between the points $A_{down}$ and $A_{up}$ situated on the surface of the sphere is placed an elastic and a first phase of the measurement consists of the ball falling through the interior of the sphere orthogonally to the elastic and sticking to it in a point $A'$ on the centerline between $A_{down}$ and $A_{up}$. The second phase of the measurement consists of the elastic breaking in one of its points randomly and uniformly, hence such that the probability of breaking in a piece of the elastic is proportional to the length of this piece. A a consequence the ball is moved upwards ending up in the point $A_{up}$ or downwards ending up in the point $A_{down}$ depending on whether the elastic breaks in the part down to $A'$ or the part up to $A'$.}
\label{ElasticSphereModelfigure}
\end{figure}

\begin{figure}[h!]
\begin{center}
\includegraphics[width=10cm]{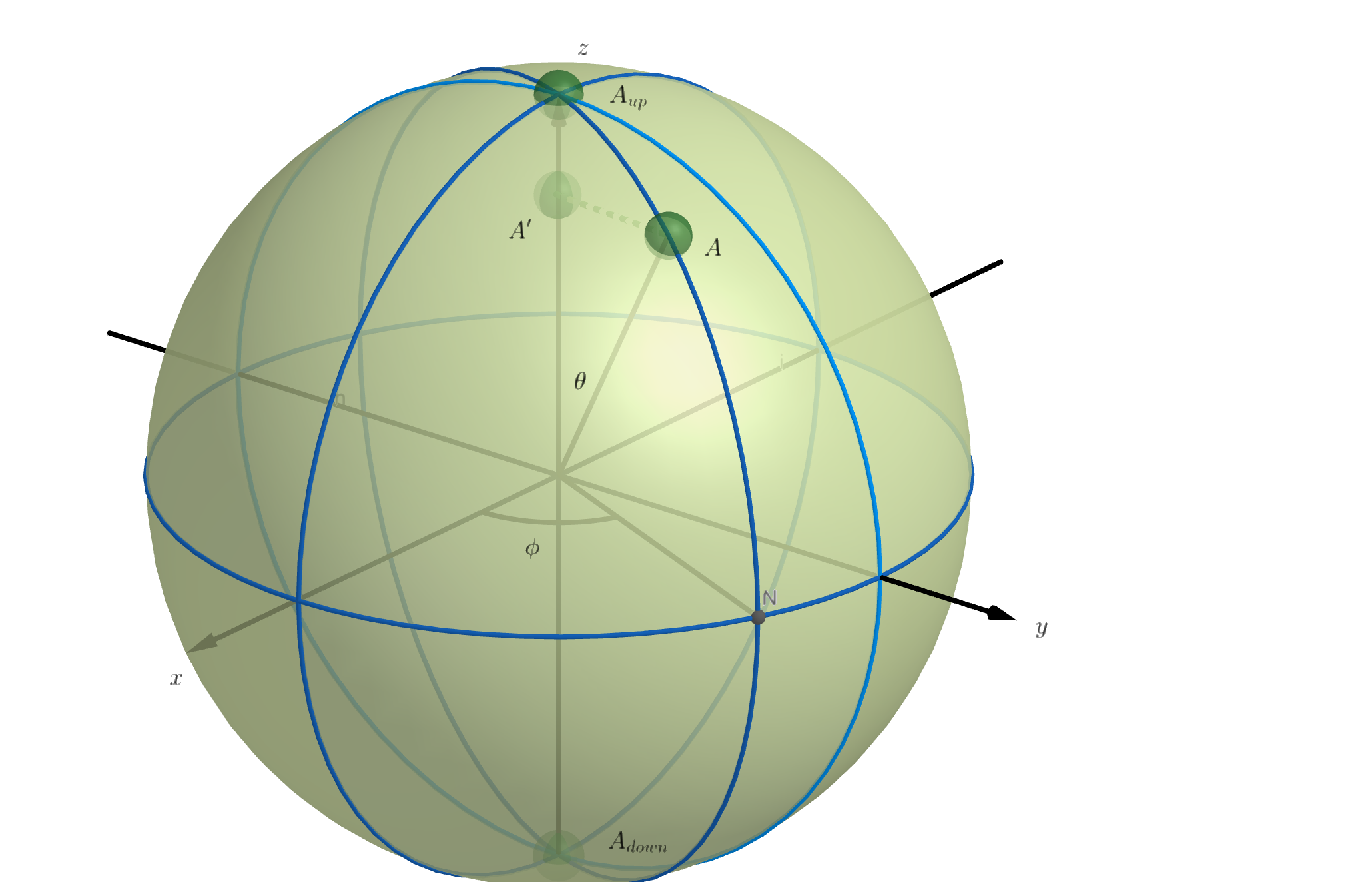}
\end{center}
\caption{A three-dimensional representation of the Extended Bloch Model. The little ball is in point $A$ with spherical coordinates $(\rho, \theta, \phi)$ and Hilbert space density operator
$D_{(\rho, \theta, \phi)}$. The process of measurement proceeds like described in detail in Figure \ref{ElasticSphereModelfigure} by the ball falling orthogonally to the elastic and sticking to it in point $A'$. Then the elastic breaks uniformly in one of its points an pulls the ball upwards to end in $A_{up}$ or downwards to end in $A_{down}$.}
\label{ElasticSphereModel3Dfigure}
\end{figure}

In Figures \ref{ElasticSphereModelfigure} and \ref{ElasticSphereModel3Dfigure} we consider the basis structure of the Bloch model for a two-dimensional quantum entity, namely the Bloch sphere, which is a sphere in a three-dimensional Euclidean space with radius equal to 1. The points on the surface of this sphere represent the pure states of the quantum entity considered. These are the states that the entity can occupy independently of being measured upon or not, thus they describe the reality of the considered entity. In the Hilbert space, these pure states are represented by unit vectors. We promised that we wanted to make the content of this article understandable for those who do not know the mathematical technicalities of the quantum formalism, and we will do our best to keep this promise. Nevertheless, we are obliged to introduce some mathematical elements that are typical of the quantum formalism. The pure states of two-dimensional quantum entity, hence a qubit, or the spin of a quantum particle with half integer spin, are represented in the extended Bloch model by the points of the surface of the Bloch sphere. 
 We will denote these points by vectors in the three-dimensional Euclidean space, indeed, the Bloch sphere is contained in this three-dimensional Euclidean space, by the so-called spherical coordinates $r$, $\theta$, and $\phi$. The distance from the center of the Bloch sphere to a point on the surface of the Bloch sphere is traditionally denoted in spherical coordinates by $r$, and called the `radial distance'. Since our Bloch sphere has radius equal to 1, for any point on the surface, we have $r=1$. The angle that the line connecting the center of the Bloch sphere with the considered point of the surface of the Bloch sphere makes with the axis of the Bloch sphere is denoted by $\theta$, and thus varies between values $0$ and $\pi$, it is equal to $0$ if the considered point coincides with the North Pole of the Bloch sphere and is called the polar angle (see Figure \ref{ElasticSphereModel3Dfigure}). Hence, it is equal to $\pi$ if it coincides with the South Pole \footnote{We denote the magnitude of angles by the unit of a radian, which makes 180 degrees equal to $\pi$ radians}.
 
A second angle, denoted by $\phi$ and called the azimuthal angle, is still needed, so that the values of $r$, $\theta$ and $\phi$
would uniquely determine each point of the three-dimensional Euclidean space. This angle $\phi$ is chosen from a vertical plane through the axis of the Bloch sphere rotating around that axis reaching the other points not in the vertical plane, for the points in the plane $\phi$ is equal to zero or $2\pi$. This means that for values of $\phi$ varying between $0$ and $2\pi$, all points of the three-dimensional Euclidean space can be reached, and hence the spherical coordinates ($r$, $\theta$, $\phi$) form complete coordination of the three-dimensional Euclidean space. For $r=1$, by varying the values of $\theta$ and $\phi$, all the points of the surface of the Bloch sphere are reached (see Figures \ref{ElasticSphereModelfigure} and \ref{ElasticSphereModel3Dfigure}). So the point $A$ has spherical coordinates $(1,\theta,\phi)$, and cartesian coordinates $(x,y,z)$, where 
 \begin{eqnarray}
x&=&\sin\theta\sin\phi \\
y&=&\sin\theta\cos\phi \\
z&=&\cos\theta
 \end{eqnarray}
for a cartesian coordinate system with its origin in the center of the sphere. 

The coordinates that we have considered so far, be they spherical $(r, \theta, \phi)$ or Cartesian $(x, y, z)$,
describe a three-dimensional Euclidean space, of which the Bloch sphere is a part. 
However, we have explicitly discussed a two-dimensional quantum entity that we want to describe using its Bloch representation, while the Cartesian and spherical coordinates describe a three-dimensional space. If one remembers the representation of the complex numbers in the Euclidean plane, a mathematics subject of high school education, then it comes as no surprise that a representation of a two-dimensional complex vector space needs three dimensions to be represented in a Euclidean real space. Indeed, the set of complex numbers, if one wishes to think of it that way, is a one-dimensional complex vector space and it needs a plane, the so-called complex plane, hence a two-dimensional space, if represented by points with coordinates that are real numbers, such as is the case by polar coordinates in the complex plane. 
Let us introduce a general vector from the two-dimensional vector space, and we immediately refer to the spherical coordinates to indicate to which point of the Bloch sphere that vector belongs. 
\begin{eqnarray} \label{statevector}
  |\theta,\phi\rangle = (\cos{\theta \over 2} e^{-i{\phi \over 2}}, \sin{\theta \over 2} e^{i{\phi \over 2}})  
\end{eqnarray}
We do not yet directly explain here why precisely this vector $|\theta, \phi \rangle$ of the two-dimensional complex Hilbert space is the one we let correspond to that point $(1, \theta, \phi)$ of the three-dimensional Euclidean space, the analysis we make in the course of this article will clarify this. 

The notation $|\theta,\phi\rangle$ is the one  introduced by Paul Dirac, one of the founding fathers of quantum mechanics, along with Werner Heisenberg and Erwin Schr\"odinger \citep{dirac1939}. Dirac considered the inner product as a fundamental operational element, the bracket, $\langle \theta, \phi|\theta, \phi\rangle$, and called the right side of this bracket, the ket, $|\theta,\phi\rangle$ and the left side, the bra $\langle \theta, \phi|$. Hence the notation $|\theta, \phi\rangle$  as the right side, thus as ket, was chosen as a representation of the pure state of the considered entity by Dirac because the inner product is taken to be `linear' in the second variable while `conjugate linear' in the first variable \footnote{This choice, by the way, is sometimes not followed by mathematicians who define the inner product in a complex Hilbert space as linear in the first variable and conjugate linear in the second.}.
In matrix notation, in the case of a two dimensional quantum entity, the ket is written as a `one column' to `two row' matrix, and the bra as a `one row' to `two column' matrix, plus a complex conjugation of the terms of the matrix for the bra.  
\begin{eqnarray}
    |\theta, \phi\rangle &=& \begin{pmatrix}
\cos{\theta \over 2} e^{-i{\phi \over 2}} \\
\sin{\theta \over 2} e^{i{\phi \over 2}}  
\end{pmatrix} \\ 
\langle \theta, \phi| &=& \begin{pmatrix}
\cos{\theta \over 2} e^{i{\phi \over 2}} &
\sin{\theta \over 2} e^{-i{\phi \over 2}}  
\end{pmatrix}
\end{eqnarray}
 We thus find by a direct matrix calculation that the ket is normalized as a state, indeed we have 
\begin{eqnarray}
    \langle \theta, \phi|\theta, \phi\rangle &=& 
    \begin{pmatrix}
\cos{\theta \over 2} e^{i{\phi \over 2}} &
\sin{\theta \over 2} e^{-i{\phi \over 2}}  
\end{pmatrix}  \begin{pmatrix}
\cos{\theta \over 2} e^{-i{\phi \over 2}} \\
\sin{\theta \over 2} e^{i{\phi \over 2}} 
\end{pmatrix} \nonumber \\ 
&=& \cos^2{\theta \over 2} + \sin^2{\theta \over 2} = 1 
\end{eqnarray}
Dirac also introduced in his bra-ket calculus the possibility of multiplying the ket by the bra in the reverse order of the bra-ket, that is, as ket-bra. Whereas the bra-ket multiplication always results in a complex number, more specifically the number 1 in the example we considered here, the reverse multiplication $ |\theta, \phi\rangle    \langle \theta, \phi|$ leads to a more complex quantity. We can calculate this more complex quantity directly in the case of the two-dimensional quantum entity we are considering, and by the representations of ket and bra by the two-by-one matrices, as we just explained. We then get 
\begin{eqnarray} \label{projector}
    |\theta, \phi\rangle    \langle \theta, \phi| 
    &=&  \begin{pmatrix}
\cos{\theta \over 2} e^{-i{\phi \over 2}} \\
\sin{\theta \over 2} e^{i{\phi \over 2}} 
\end{pmatrix}  \begin{pmatrix}
\cos{\theta \over 2} e^{i{\phi \over 2}} &
\sin{\theta \over 2} e^{-i{\phi \over 2}}  
\end{pmatrix} \nonumber \\ 
&=&  \begin{pmatrix}
\cos^2{\theta \over 2} & \cos{\theta \over 2}\sin{\theta \over 2}e^{-i\phi} \\
\cos{\theta \over 2}\sin{\theta \over 2}e^{i\phi} & \sin^2{\theta \over 2}
\end{pmatrix} \label{theta0} 
\end{eqnarray}
 a `two-by-two' matrix. This two-by-two matrix represents an operator within the quantum Hilbert space formalism called an orthogonal projection, but before we continue on that, we wish to specify using the extended Bloch model what happens during a quantum measurement. 

Within the extended Bloch model, the measurement is represented by a piece of elastic stretched on the centerline between two diametrically opposed points of the sphere, $A_{down}$ and $A_{up}$ (see Figures \ref{ElasticSphereModelfigure} and \ref{ElasticSphereModel3Dfigure}). On the surface of the sphere is a small ball that sticks to the sphere at a point $A$. This ball in a point represents the state in which the quantum entity is. The measurement then occurs as follows. First, the ball falls orthogonally on the elastic and sticks to it at the point $A'$ where this orthogonal fall brings it. Then the elastic breaks randomly at one of its points. If the breaking point is below the point $A'$, then the unbroken part of the elastic pulls the ball up so that it ends up in the point $A_{up}$. However, if the breaking point is above the point $A'$, then the ball is pulled down by the unbroken part of the elastic and ends up in the point $A_{down}$. Thus, the whole measurement results in the ball from point $A$ ending up at one of two points $A_{up}$ or $A_{down}$. 

Before we proceed with the description of the measurement we note that of course there is not really an elastic present in a quantum measurement, so we owe some explanation of the reason for introducing such an elastic in order to represent the measurement as visually as possible within the mathematical framework of the Bloch sphere. In the further course of the stepwise development of both the algebraic and the geometric representation we put forward, it will become clear that the purpose of introducing the elastic is to offer a pictorial model of the so-called quantum collapse. But also the quantum probabilities are easily derived from the breaking of the elastic as we will show. Furthermore, the process of decoherence also finds a simple geometric description through it. 

Let us begin by showing how the quantum probabilities are simply derived from the geometry of the Bloch configuration and a hypothesis of uniformity of the elastic. So, we suppose that the elastic possesses a uniform breaking pattern, which means that the point at which the elastic breaks randomly will lie in a certain interval of the elastic with a probability proportional to the length of that interval. We can then calculate the probabilities $P(A \mapsto A_{up})$ and $P(A \mapsto A_{down})$ with which the measurement will carry the ball from $A$ to $A_{up}$ or to $A_{down}$, and in this way will change the state of the quantum entity, from the simple geometry of the configuration. Note indeed that we have that the length $L_2$ of the elastic under the point $A'$ equals $1+\cos\theta$ and the length $L_1$ of the elastic above the point $A'$ equals $1-\cos\theta$, which hence, taking into account the uniform nature of the breaking pattern of the elastic, gives us
\begin{eqnarray} \label{qubit01}
P(A \mapsto A_{up}) &=& {1+\cos\theta \over 2} =  \cos^2 {\theta \over 2} \\ \label{qubit02}
P(A \mapsto A_{down}) &=& {1-\cos\theta \over 2} = \sin^2 {\theta \over 2}
\end{eqnarray}
The probabilities (\ref{qubit01}) and (\ref{qubit02}) match the quantum probabilities of the spin of a spin 1/2 quantum particle or a qubit. The ball at point $A$ is then in a state for the spin of an angle $\theta$ with the $z$-axis, and the elastic lies on this $z$-axis. 

The part of this quantum measurement that takes place at the so-called pure states is described in a complete way in the quantum Hilbert space formalism. The complex Hilbert space in this case is two-dimensional, with basis vectors $|0, 0 \rangle$ and $|\pi, 0\rangle$, which in the Bloch representation correspond to points on the Bloch sphere located at the North and South Poles, respectively. In Hilbert space these basis vectors, and we use now their notation as one column and two row matrices, are 
\begin{eqnarray}
    |0, 0\rangle = \begin{pmatrix}
\cos{\theta \over 2} e^{-i{\phi \over 2}} \\
\sin{\theta \over 2} e^{i{\phi \over 2}} 
\end{pmatrix}_{(\theta=0,\phi=0)} = \begin{pmatrix}
1 \\
0
\end{pmatrix} \\ 
|\pi, 0\rangle = \begin{pmatrix}
\cos{\theta \over 2} e^{-i{\phi \over 2}} \\
\sin{\theta \over 2} e^{i{\phi \over 2}} 
\end{pmatrix}_{(\theta=\pi,\phi=0)} = \begin{pmatrix}
0 \\
1 
\end{pmatrix}
\end{eqnarray}
A general pure state $|\theta, \phi\rangle$, in this Hilbert space formalism, is a normalized linear combination, called superposition in the quantum jargon, of these basis vectors
\begin{eqnarray}
    && |\theta, \phi\rangle = \begin{pmatrix}
\cos{\theta \over 2} e^{-i{\phi \over 2}} \\
\sin{\theta \over 2} e^{i{\phi \over 2}} 
\end{pmatrix} =  a \begin{pmatrix}
1 \\
0
\end{pmatrix} + b \begin{pmatrix}
0 \\
1 
\end{pmatrix} \\ 
&& {\rm with}\ a = \cos{\theta \over 2} e^{-i{\phi \over 2}} {\rm and}\ b = \sin{\theta \over 2} e^{i{\phi \over 2}} {\rm and}\ |a|^2 + |b|^2 = 1
\end{eqnarray}
with superposition coefficients $a$ and $b$.

We now know the quantum probabilities, namely $\cos^2{\theta \over 2}$ and $\sin^2{\theta \over 2}$, for the state $|\theta, \phi \rangle$ to collapse as a result of a measurement to the state $|0, 0 \rangle$, i. e.  the North Pole of the Bloch sphere or the state $|0, \pi\rangle$, i. e. the South Pole of the Bloch sphere, respectively. That information gives us a way to determine the mathematical form in the complex two-dimensional Hilbert space that the state $|\theta, \phi\rangle$ possesses. Indeed, let us write the most general mathematical form for this state $|\theta, \phi\rangle$, namely 
\begin{eqnarray}
    |\theta, \phi\rangle = a |0, 0\rangle + b |\pi, 0\rangle
\end{eqnarray}
where $a$ and $b$ are two complex numbers such that $|a|^2+|b|^2 = 1$, we just apply the superposition principle here, one of the basic principles of quantum theory. And then we must have 
\begin{eqnarray}
    |a|^2 = \cos^2{\theta \over 2} \quad {\rm and} \quad |b|^2 = \sin^2{\theta \over 2}
\end{eqnarray}
We can solve these equations and should not forget that $a$ and $b$ are complex numbers. A straightforward solution is $a = \cos{\theta \over 2}$ and $b = \sin{\theta \over 2}$, but that is not the general solution. Note that a dependence of a solution on $\phi$ must vanish in $|a|^2$ and $|b|^2$, which means that this dependence on $\phi$ is in the form of a phase factor $e^{i\phi}$, since such a phase factor disappears if the absolute value of the complex numbers is calculated. This is how we arrive at the general solution $a = \cos{\theta \over 2} e^{-i{\phi \over 2}}$ and $b = \sin{\theta \over 2} e^{i{\phi \over 2}} $, which we introduced in (\ref{statevector}), noting that without loss of generality we can multiply with a phase factor such that this specific form arises \footnote{Another choice often found in textbooks is $a = \cos{\theta \over 2}$ and $b = \sin{\theta \over 2}e^{i\phi}$, which equals the solution we use in this article except a phase of $e^{-i{\phi \over 2}}$.}. 
The role of complex numbers is very fundamental to quantum mechanics and far from fully understood. Thus, it can be shown that the probability structure present in a two-dimensional quantum entity cannot be obtained in its completeness within a probability model that does not use the underlying structure of a complex vector space \citep{aerts1986}. 

What we are particularly interested in for our present article is the fine structure of the mechanism in the measurement in the extended Bloch model shown in Figures \ref{ElasticSphereModelfigure} and \ref{ElasticSphereModel3Dfigure}. Before we proceed, we wish to note the following. The example depicts a two-dimensional quantum entity and a measurement with two final states $A_{up}$ and $A_{down}$, hence two outcomes. However, in later years, it was shown that a similar model can be built for a $n$-dimensional quantum entity and a measurement with an arbitrary number $n$ of final states and thus an arbitrary number $n$ of outcomes \citep{aertssassolidebianchi2014}. 
We will not explicitly describe these higher-dimensional extended Bloch models in the present article, but mention that the simple geometric properties of the configurations lead to exactly the quantum probabilities in a completely similar way than this is the case for this two-dimensional quantum entity. The details of the analysis that we will now make for the two-dimensional quantum entity can also be made for the higher-dimensional quantum entities in a similar way. Of course, finding exact quantum probabilities in these measurement models also depends on a certain symmetry that we introduced at the measurement level, namely that the elastic breaks with a randomness that is `uniformly' distributed. If the elastic does not possess this symmetry property, we will still find probabilities that do not fit a classical probability model, but they will not be perfect quantum probabilities, e.g. a complex Hilbert space model will not be able to represent them, and they will thus be rather quantum-like. In this case, a more general formalism than standard quantum mechanics can be used, which is not based on vectors representing states \citep{aerts1982,aerts1992,gudderzanghi1984}. 

To verify that a quantum measurement gives rise to the mechanism of categorical perception, we must be able to express for states how much they differ from each other. But before we get to that, we must identify which states play the role of stimuli and which states play the role of percepts in a quantum measurement process. The stimuli are represented by the states of the quantum entity independent of any measurement that would be going on, thus, they are what in the quantum jargon are called the pure states. In the Hilbert space, these pure states are represented by unit vectors. In the Bloch representation, it is the points of the surface of the Bloch sphere that represent the pure states of the considered quantum entity. 
 
In what happens during a measurement as we represent it in the extended Bloch model, not only the places corresponding to points on the surface of the Bloch sphere play a role, but also places corresponding to points of the interior of the Bloch sphere are important, for example the point $A'$ where the ball lands and sticks to the elastic after falling orthogonally on it. The points over which the elastic is stretched are also in the interior of the Bloch sphere. Besides the vector space formalism of quantum mechanics, where the main role is played by the complex Hilbert space, which is a vector space over the complex numbers, there also exists a density matrix formalism of quantum mechanics. However, both formalism have a fundamentally different mathematical structure, the Hilbert space is a vector space, as we already mentioned, while the set of density matrices is a convex space. Convex combinations of density matrices  give another density matrix, while normalized linear combinations of unit vectors give another unit vector. Also, both structures can live together on an underlying mathematical entity, e.g. density matrices are found in the Hilbert space formalism as density operators. Both are also found in the Bloch representation, the unit vectors of Hilbert space as the points of the surface of the Bloch sphere, and the density matrices as the points of the interior of the Bloch sphere. 
Let us introduce more systematically these mathematical structures by writing the different vectors with respect to the canonical basis. 
\begin{eqnarray} 
  |1, 0\rangle = (1, 0) \\
  |0, 1\rangle = (0, 1) 
\end{eqnarray} 
We have 
\begin{eqnarray} \label{statevector2}
  |\theta,\phi\rangle = (\cos{\theta \over 2} e^{-i{\phi \over 2}}, \sin{\theta \over 2} e^{i{\phi \over 2}})  
\end{eqnarray} 
We already mentioned the density matrix formalism as also a quantum mechanical formalism. The points of the interior of the Bloch sphere correspond to such density matrices of this quantum formalism. We will denote a density operator corresponding to the point with spherical coordinates $(r, \theta, \phi)$, $\rho \in [0,1]$, $\theta \in [0,\pi]$, $\phi \in [0, 2\pi]$ as $D_{(r, \theta, \phi)}$. Let us use some of the known properties to calculate the density matrix for some of the points of the Bloch sphere and of its interior. 
We already introduced the density matrix corresponding to a pure state in (\ref{projector}), hence in the notation we introduced this is $D_{(1,\theta,\phi)}$. Let us calculate the density matrices of the North and South Poles of the Bloch sphere, or of the points $A_{up}$ and  $A_{down}$. We have 
\begin{eqnarray}
D_{(1,0,\phi)} &=& \begin{pmatrix}
\cos^2{\theta \over 2} & \cos{\theta \over 2}\sin{\theta \over 2}e^{-i\phi} \\
\cos{\theta \over 2}\sin{\theta \over 2}e^{i\phi} & \sin^2{\theta \over 2}
\end{pmatrix}_{\theta=0,\phi} 
= \begin{pmatrix} 
1 & 0 \\
0 & 0 
\end{pmatrix}
\\ \label{thetapi}
D_{(1,\pi,\phi)} &=&\begin{pmatrix}
\cos^2{\theta \over 2} & \cos{\theta \over 2}\sin{\theta \over 2}e^{-i\phi} \\
\cos{\theta \over 2}\sin{\theta \over 2}e^{i\phi} & \sin^2{\theta \over 2} 
\end{pmatrix}_{\theta=0,\phi} 
= \begin{pmatrix}
0 & 0 \\
0 & 1
\end{pmatrix}
\end{eqnarray}
Let us now calculate the density operator $D_{A'}$ representing the state of the entity when it is in the point $A'$ as in Figure \ref{ElasticSphereModel3Dfigure}. We remark that $A'$ lies on the line between $A_{down}$ and $A_{up}$ sticking on the elastic which is stretched between $A_{down}$ and $A_{up}$ on this line. Making use of a general property of the set of all density operators, namely that it is a set closed by convex combination, we know that $D_{A'}$ is a convex combination of $D_{(1,\pi,\phi)}$ and $D_{(1,0,\phi)}$, which gives
\begin{eqnarray} \label{convex01}
D_{A'} = \lambda \begin{pmatrix}
0 & 0 \\
0 & 1
\end{pmatrix}
+ (1-\lambda) \begin{pmatrix} 
1 & 0 \\
0 & 0 
\end{pmatrix}
= \begin{pmatrix}
1-\lambda & 0 \\
0 & \lambda
\end{pmatrix}
\end{eqnarray}
for $\lambda \in [0, 1]$. From Figure \ref{ElasticSphereModel3Dfigure}, given that $A'$ is obtained by projecting orthogonally to the line between $A_{down}$ and $A_{up}$, we have that $A'$ lies on the line between $A$ and the point with spherical coordinates $(1,\theta,\phi+\pi)$, to which corresponds the density operator
\begin{eqnarray}
D_{(1,\theta,\phi+\pi)} &=& 
\begin{pmatrix}
\cos^2{\theta \over 2} & \cos{\theta \over 2}\sin{\theta \over 2}e^{-i(\phi+\pi)} \\
\cos{\theta \over 2}\sin{\theta \over 2}e^{i(\phi+\pi)} & \sin^2{\theta \over 2}
\end{pmatrix} \nonumber \\
&=&
\begin{pmatrix}
\cos^2{\theta \over 2} & -\cos{\theta \over 2}\sin{\theta \over 2}e^{-i\phi} \\
-\cos{\theta \over 2}\sin{\theta \over 2}e^{i\phi} & \sin^2{\theta \over 2}
\end{pmatrix}
\end{eqnarray}
This means that we have
\begin{eqnarray} \label{convex02}
D_{A'} = \mu D_{(1,\theta,\phi)} + (1-\mu) D_{(1,\theta,\phi+\pi)}
\end{eqnarray}
for $\mu \in [0, 1]$. From (\ref{convex01}) and (\ref{convex02}) follows that we must have
\begin{eqnarray}
&&\mu \cos{\theta \over 2}\sin{\theta \over 2}e^{-i\phi} - (1-\mu) \cos{\theta \over 2}\sin{\theta \over 2}e^{-i\phi} =0 \nonumber \\
&&\Leftrightarrow 
\mu - (1-\mu) = 0 \nonumber \\
&& \Leftrightarrow
\mu = {1 \over 2} 
\end{eqnarray} 
and 
\begin{eqnarray}
\lambda = \sin^2{\theta \over 2} 
\end{eqnarray} 
This gives us
\begin{eqnarray} \label{densitystate}
D_{A'} &=& \begin{pmatrix}
\cos^2{\theta \over 2} & 0 \\
0 & \sin^2{\theta \over 2}
\end{pmatrix} =
\cos^2{\theta \over 2}
\begin{pmatrix}
1 & 0 \\
0 & 0
\end{pmatrix}
+ \sin^2{\theta \over 2}
\begin{pmatrix}
0 & 0 \\
0 & 1
\end{pmatrix} \nonumber \\
&=& \cos^2{\theta \over 2} D_{A_{up}} + \sin^2{\theta \over 2} D_{A_{down}}
\end{eqnarray}
Let us note that it follows from (\ref{convex02}) that we also know the density matrix of each point lying on the line between the two points of the surface of the Bloch sphere, the point $A$ with coordinates $(1, \theta, \phi)$ and the point with coordinates $(1, \theta, \phi + \pi)$, which is mirrored with respect to the North-South axis of the Bloch sphere. For a value of $\mu$ ranging from -1 to +1, the density matrix is given by 
\begin{eqnarray}
D_{A',\mu} &=&  \mu D_{(1,\theta,\phi)} + (1-\mu) D_{(1,\theta,\phi+\pi)} \nonumber \\
&=& \begin{pmatrix}
\cos^2{\theta \over 2} & \mu \cos{\theta \over 2}\sin{\theta \over 2}e^{-i(\phi+\pi)} \\
\mu \cos{\theta \over 2}\sin{\theta \over 2}e^{i(\phi+\pi)} & \sin^2{\theta \over 2}
\end{pmatrix} 
\end{eqnarray}
the density matrix of the point corresponding to the value of $\mu$ on the line between the points $(1, \theta, \phi + \pi)$ and $(1, \theta, \phi)$. The parameter $\mu$ occurs only in the diagonal terms of the density matrix, and the orthogonal falling of the ball on the elastic takes place on the line connecting the two points $(1, \theta, \phi + \pi)$ and $(1, \theta, \phi)$, and that orthogonal falling occurs when $\mu$ goes from 1 to 0. 
In the jargon of quantum mechanics, the decreasing of the non-diagonal terms of the density matrix is named `decoherence,' and is described as a phenomenon of decreasing quantum nature and increasing classical nature. 

There is another way to consider the vanishing of quantum coherence, which, since in this article we are investigating how the mechanism of categorical perception is present in the quantum measurement process, is important to mention and analyze. More concretely, the density matrix corresponding to the density state described by the point $A'$ of the interior of the Bloch sphere is also found as the trace class matrix with respect to the measurement device, when we describe the measurement process as an entanglement coupling of the measurement device with the quantum entity. 

In this coupling, both lose their identity as a separate entity for a brief period, the time in which the measurement takes place, and merge into the larger system consisting of two subsystems, one being the measuring device and the other the quantum entity. In that period, that larger system is in an entangled state, which is a pure state, described by a unit vector of a higher-dimensional Hilbert space. Let us work out the explicit mathematical model of this situation, where we describe the measuring device by a two-dimensional Hilbert space with basis vectors $|up\rangle$ and $|down\rangle$, and the quantum entity likewise by a two-dimensional Hilbert space, the one we already introduced. The larger entity, consisting of the measuring device and the quantum entity, is then described in a four-dimensional Hilbert space, which is the tensor product of the two Hilbert spaces of the subsystems. An orthonormal basis of this tensor product is given by the set of vectors $\{|up\rangle \otimes |0, \phi \rangle, |up\rangle \otimes |\pi, \phi \rangle, |down\rangle \otimes |0, \phi \rangle, |down\rangle \otimes |\pi, \phi \rangle \}$
where
$\{ |up\rangle, |down\rangle \}$ is an orthonormal basis of the measuring apparatus and $\{|0, \phi \rangle, |\pi, |\phi \rangle   \}$ is an orthonormal basis of the quantum entity considered.
Let us call $|\psi \rangle_e$ an entangled state of this joint entity consisting of the measuring apparatus and the quantum entity considered, then we can write
\begin{eqnarray}
| \psi_e \rangle &=& a_{11} |up\rangle \otimes |0, \phi \rangle + a_{10} |up\rangle \otimes |\pi, \phi \rangle + 
a_{01}  |down\rangle \otimes |0, \phi \rangle 
+ a_{00} |down\rangle \otimes |\pi, \phi \rangle 
\end{eqnarray}
where $\{a_{11}, a_{10}, a_{01}, a_{00} \}$ are the amplitudes of this general entangled state. However, if we write such a general entangled state for the composite entity of measuring device and quantum entity, we have not yet expressed that we are dealing with a very special situation, namely that of a measurement. Specific to this situation is that the outcome state of the measuring device $|up\rangle$ should only coincide with the state $|0, \phi \rangle$ of the quantum entity and that the outcome state of the measuring device $|down\rangle$ should only coincide with the state $|\pi, \phi \rangle$ of the quantum entity, which means that two amplitudes, namely $a_{01}$ and $a_{10}$ are always equal to zero. Also the two other amplitudes, $a_{11}$ and $a_{00}$, of the entangled state of the joint entity measurement device and quantum entity are determined, since we know the Born probabilities, $\cos^2{\theta \over 2}$ and $\sin^2{\theta \over 2}$ as limits of the relative frequencies of repeated measurements. Hence, this gives us 
 $a_{11} =  \cos{\theta \over 2}$ and $a_{00} = \sin{\theta \over 2}$ as possible choices for the remaining non zero amplitudes. Let us write now the entangled state $|\psi_e\rangle$ of the joint entity consisting of the measuring device and the quantum entity considered. 
\begin{eqnarray}
    |\psi_e\rangle = \cos{\theta \over 2}|up\rangle \otimes |0, \phi \rangle + \sin{\theta \over 2}|down\rangle \otimes |\pi, \phi \rangle
\end{eqnarray}
We emphasize again that this unit vector of the four-dimensional Hilbert space that is the tensor product represents a pure state of the joint entity consisting of the measuring device and the considered quantum entity. Since it is an entangled state neither the measuring device nor the considered quantum system are in a pure state, so it is a state in which this joint entity is in for a brief period of time. The measurement has not yet ended, by the way, because no outcome has been obtained, both outcomes up and down are still possible. Also the quantum entity is not in a pure state in this period of time in which the measurement is taking place. The quantum formalism gives us the technique to calculate the density states in which both, the measuring device and the quantum entity are, while the measurement is taking place. Let us make this calculation. 

Hence, let us first calculate the density state corresponding to the pure state $|\psi_e\rangle$.
We have  
\begin{eqnarray}
| \psi_e \rangle\langle \psi_e| &=& a_{11} a^*_{11} |up\rangle \langle up | \otimes |0, \phi \rangle \langle 0, \phi |
+ a_{00} a^*_{11} |down\rangle  \langle up | \otimes |\pi, \phi \rangle \langle 0, \phi | \nonumber \\ 
&& + a_{11} a^*_{00} |up\rangle \langle down| \otimes |0, \phi \rangle \langle \pi, \phi | + a_{00}  a^*_{00} |down\rangle  \langle down| \otimes |\pi, \phi \rangle \langle \pi, \phi |
\end{eqnarray}
This density state is a pure state of the joint entity of the measuring device and the considered quantum entity. The partial trace density states give the states in which the sub-entities, that is, the measuring device and the quantum entity are when the joint entity is in this pure state. Let us calculate these partial trace states. We have for the partial trace density state $\rho^{md}$ of the measuring device 
\begin{eqnarray}
\rho^{md} &=& a_{11} a^*_{11} \langle up| up \rangle \otimes |0, \phi \rangle \langle 0, \phi |
+ a_{00} a^*_{11} \langle down |  up \rangle  \otimes |\pi, \phi \rangle \langle 0, \phi | \nonumber \\ 
&& + a_{11} a^*_{00} |
\langle up | down \rangle \otimes |0, \phi \rangle \langle \pi, \phi | + a_{00}  a^*_{00} \langle 
down | down \rangle \otimes |\pi, \phi \rangle \langle \pi, \phi | \nonumber \\
&=& |a_{11}|^2 |0, \phi \rangle \langle 0, \phi | + |a_{00}|^2 |\pi, \phi \rangle \langle \pi, \phi | \nonumber \\
&=& \cos^2{\theta \over 2} \begin{pmatrix}
    1 & 0 \\
    0 & 0
\end{pmatrix} + 
\sin^2{\theta \over 2} \begin{pmatrix}
    0 & 0 \\
    0 & 1
\end{pmatrix} \nonumber \\
&=& \begin{pmatrix}
\cos^2{\theta \over 2} & 0 \\
0 & \sin^2{\theta \over 2} 
\end{pmatrix} 
\end{eqnarray}
Note that this density state $\rho^{md}$ of the measuring device is equal to the density state corresponding to the point $A'$ (see \ref{densitystate}). 

A similar calculation shows that the partial trace density state $\rho^{qe}$ corresponding to the quantum entity considered is also equal to the density state corresponding to $A'$.

We will use the detailed way in which we have described the quantum measurement to answer some questions that are often asked. The measuring device is a classical entity, so how come in our elaborated description we use the notion of superposition state for this classical entity? Well, the measuring device is indeed a classical entity when studied as an entity, but it is also a special classical entity, namely, quantum entities can be measured with it. In the period in which such a measurement is in progress and the result is not yet obtained, as it were, the measuring device for a short period becomes a quantum entity which in this period where the measurement takes place merges into a sub-entity of the joint entity consisting of the measuring device and the quantum entity to be measured. This joint entity is then in an entangled state of both sub-entities. The measurement, however, proceeds so that one of the two outcomes is obtained and this occurs because the joint entity breaks up into the two original entities, measurement device and quantum entity, where both end up in a specific state, the measurement device in one of the two possible outcome states and the quantum entity in the eigen-state belonging to this outcome state. 

\bigskip 
\noindent 
{\bf Summery} 

\medskip 
\noindent 
Let us briefly summarize the elements of a quantum measurement and its Bloch sphere representation. 
 Each point $(r, \theta, \phi)$,  $r \in [0,1]$, $\theta \in [0,\pi]$, $\phi \in [0,2\pi]$, within the sphere shown in Figure \ref{ElasticSphereModel3Dfigure}, corresponds to a density operator $D_{(r,\theta,\phi)}$, and the points on the surface of the sphere, that is, $r=1$, correspond to the pure states of the quantum entity. The points of the interior of the sphere correspond to density states of the quantum entity. With a measurement corresponds a convex subspace of the interior of the sphere, in the case of the measurement with possible final states after the measurement, $A_{up}$ and $A_{down}$, this convex subspace is given by the line segment from $(1,\pi,\phi)$ to $(1,0,\phi)$, where the elastic is stretched. But all lines passing through the center of the sphere in Figure \ref{ElasticSphereModel3Dfigure} harbor such a measurement. The points on the line all correspond to a density state, except for the end points, which are indeed part of the surface of the sphere and thus correspond to a pure state. 
 
 When a measurement begins, on an entity that is in a pure state, such as this one in the point $A$, an expectation is allowed to play a role of what this state in the point $A$ means to the measuring apparatus located on a line between two diametric points of the surface of the sphere, such as the points $A_{down}$ and $A_{up}$. 
 
That is the meaning of the orthogonal projection which brings the state in $A$, to a state in $A'$. Once the state has become the one corresponding to the point $A'$, the entity is in a density state. The transformation from the state corresponding to $A$ to the state corresponding to $A'$ can be read on the matrix representation, that is, it is the non-diagonal terms of the density state matrix that disappear to arrive at the state corresponding to $A'$. 

This change from $A$ to $A'$, described mathematically at the level of the density formalism by decreasing the non-diagonal terms of the density matrix, and described geometrically in the Bloch sphere by the orthogonal dropping of the little ball on the elastic sticking to it, is called `decoherence'. It describes a process of `decrease of quantum coherence'. As long as the density matrix corresponds to that of a pure state, that is, as long as the state is at a point on the surface of the Bloch sphere, a situation of complete quantum coherence prevails. From the moment the non-diagonal terms of the density matrix begin to decrease, that is, the state begins to move on the line in the interior of the Bloch sphere orthogonal to the elastic, the quantum coherence begins to decrease, to disappear completely at the moment the non-diagonal terms become zero, which is the moment the state arrives in $A'$. The original quantum coherence of the state has then completely disappeared, but note that the measurement is not yet at an end. We emphasize this because erroneously within an interpretation of quantum mechanics called the `decoherence interpretation', the impression is often given that after this process of decoherence the measurement has ended. It is then also incorrectly claimed that the interpretation does not require a `collapse of the wave function' aspect of the measurement. A look at what takes place mathematically and what takes place geometrically in the Bloch sphere easily shows the error of these claims from the decoherence interpretation. In fact, the collapse part of the measurement must still occur after the decoherence process, and it consists of the density matrix obtained corresponding to point $A'$ transforming into one of the pure states corresponding to $A_{up}$ or $A_{down}$. We represent this collapse event geometrically by the breaking of the elastic. Note, by the way, that this restores quantum coherence after the measurement, albeit in a different state. 

This dynamics of the quantum measurement can also be described, from a more general perspective, by an initial stage where the measuring device and the quantum entity merge into a larger entity consisting of the measuring device and the quantum entity and find themselves for a short time in a pure entangled state of this joint entity. It can be shown that the trace class operator of this entangled state is given by the density matrix corresponding to $A'$, and therefore this merging into this joint entity corresponds to the process of decoherence. The collapse, in this more general approach to the quantum measurement, then corresponds to the breaking apart of this joint entity again, with the measuring device yielding one of the outcomes of the measurement, while the quantum entity ends up in the eigenstate belonging to that outcome.

\section{Categorical Perception in Quantum Measurement \label{categoricalperceptionquantummeasurement}}

In this section, we want to show that a quantum measurement structurally incorporates the warping of the categorical perception mechanism. If we want to give an expression to the mechanism of categorical perception, we must be able to estimate and/or measure distances between both pure states, the stimuli, located in points on the surface of the Bloch sphere, and density states, the percepts, located in the points of a centerline of the Bloch sphere. Using these distances, the differences between the stimuli and the differences between the percepts can be expressed quantitatively. However, we must be careful in choosing how we will measure distances in the Bloch-sphere representation, since there is no unique obvious metric on the whole set of quantum states. 

\begin{figure}[h!]
\begin{center}
\includegraphics[width=7cm]{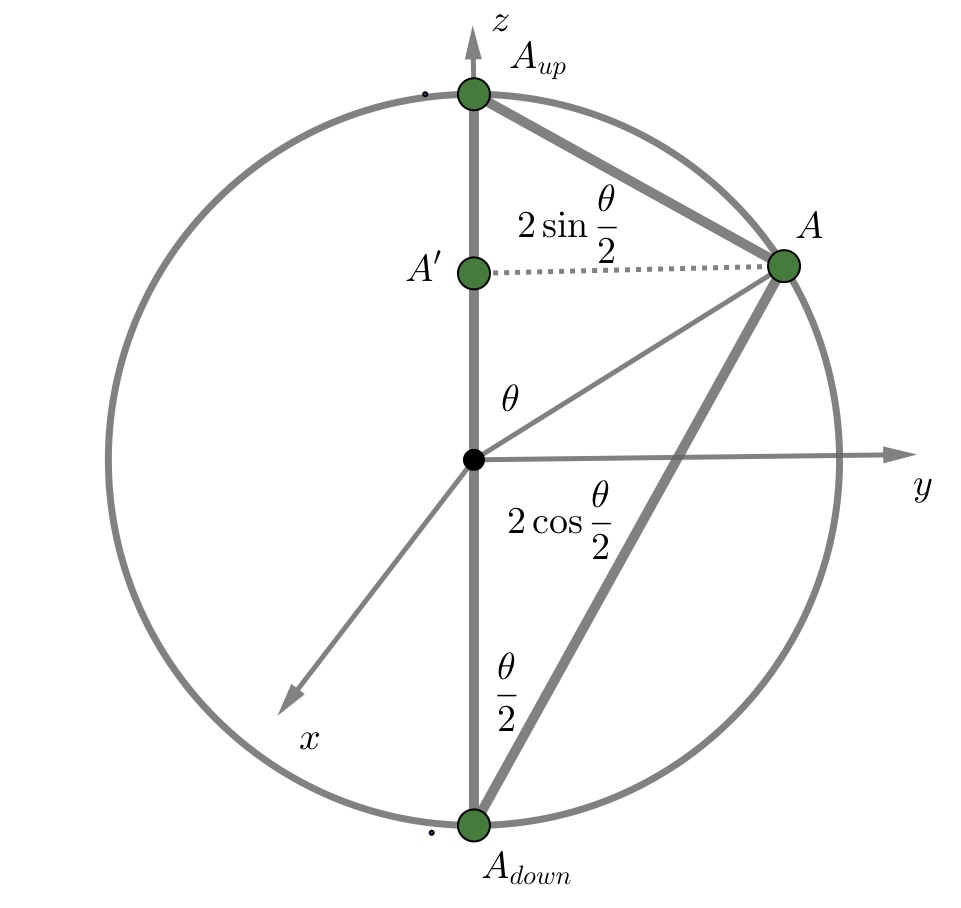}
\end{center}
\caption{Using the simple geometry of the Bloch sphere, we can recover the trace distance we calculated in (\ref{Pythagorasup}) and (\ref{Pythagorasdown}) as illustrated in this figure. The triangle $A_{up}$, $A$, $A_{down}$ is rectangular which makes its angle in $A_{down}$ equal to ${\theta \over 2}$, and consequently, knowing that the distance between $A_{up}$ and $A_{down}$ is equal to $2$, the figure shows us that the distance between $A_{up}$ and $A$ is equal to $2\sin{\theta \over 2}$, and the distance between $A_{down}$ and $A$ is equal to $2\cos{\theta \over 2}$. Since the trace distance is equal to half the Euclidean distance for the Bloch sphere of a qubit, we have derived the results of (\ref{Pythagorasup}) and (\ref{Pythagorasdown}) from the geometry of the Bloch sphere.}
\label{PythagorasBloch02ref}
\end{figure}

One of the well-studied partial metrics on the set of quantum states is the trace distance 
\begin{eqnarray}
T(D_1, D_2) &=& {1 \over 2} Tr \left [\sqrt{(D_1-D_2)^*(D_1-D_2)} \right ]
\end{eqnarray}  
It can be shown that for a two-dimensional quantum entity, i.e., a qubit, modeled with the Bloch sphere, the trace distance is equal to half the Euclidean distance in the three-dimensional Euclidean space of which the Bloch sphere is a part. For two pure states $|\psi_1\rangle$ and $|\psi_2\rangle$, hence represented by two points of the surface of the Bloch sphere, it can be shown that the trace distance is given by
\begin{eqnarray} \label{tracepurestates}
   T(|\psi_1 \rangle \langle \psi_1 |, |\psi_2 \rangle \langle \psi_2 |) &=& \sqrt{1 - |\langle \psi_1 | \psi_2 \rangle |^2}
\end{eqnarray}
That means, for example, that the trace distances between the quantum state where $A$ is and the North and South Pole of the Bloch sphere are given by
\begin{eqnarray} \label{Pythagorasup}
   T(D_{(1, \theta, \phi)}, D_{(1, 0, 0)}) &=& 
    \sqrt{1 - |\langle \theta, \phi | 0, 0 \rangle |^2} = \sqrt{1 - |\langle (\cos{\theta \over 2} e^{-i{\phi \over 2}}, \sin{\theta \over 2} e^{i{\phi \over 2}})|  (1, 0) \rangle |^2} \nonumber \\ 
   &=& \sqrt{1 - \cos^2{\theta \over 2}} = \sin{\theta \over 2} 
\end{eqnarray}
\begin{eqnarray}
   \label{Pythagorasdown}
   T(D_{(1, \theta, \phi)}, D_{(1, \pi , 0)}) &=& 
    \sqrt{1 - |\langle \theta, \phi | \pi , 0 \rangle |^2} = \sqrt{1 - |\langle (\cos{\theta \over 2} e^{-i{\phi \over 2}}, \sin{\theta \over 2} e^{i{\phi \over 2}})|  (0, 1) \rangle |^2} \nonumber \\ 
   &=& \sqrt{1 - \sin^2{\theta \over 2}} = \cos{\theta \over 2} 
\end{eqnarray}
We can now see interesting connections using the simple geometry of the Bloch representation, let us consider Figure \ref{PythagorasBloch02ref}. It represents the Bloch representation with $\phi = {\pi \over 2}$, so $A$ is laying on the $zy$ plane. The triangle $A_{up}$, $A$, $A_{down}$ is a triangle with a right angle in $A$. The hypotenuse of this triangle is the center line that connects $A_{up}$ to $A_{down}$ and has a length equal to $2$. 

The angle in $A_{down}$ of the triangle is equal to half the angle $\theta$, so the length of the line segment connecting $A_{up}$ to $A$ is equal to $2\sin{\theta \over 2}$. Similarly, the figure shows us that the length of the line segment connecting $A$ and $A_{down}$ is given by $2\cos{\theta \over 2}$. This agrees with the calculation of the trace distance we made for the states connected to the corresponding points of the Bloch sphere, as given in (\ref{Pythagorasup}) and (\ref{Pythagorasdown}), reminding us that for a qubit the trace distance is equal to half the Euclidean distance. This reasoning and additional calculations, as well as the general expression for the trace distance between pure states, make it clear that the trace distance is not the metric to be used between pure states. Indeed, any effect due to quantum coherence disappears when using this metric, since in the general expression (\ref{tracepurestates}) the only reference to the states is the inner product between the two states, and hence traces of quantum coherence effects are no longer present. 

There is also a purely geometric way by which we can see that the trace distance is unsuitable for measuring distances between pure states. Consider any two pure states, which gives us two points of the surface of the Bloch sphere. If we connect these two points with a straight line, we see that the points of the line different from these two points always lie in the interior of the Bloch sphere and thus correspond to density states. It is clear that defining a distance associated with a specific geometric entity, the straight line in the three-dimensional Euclidean space to which the Bloch sphere belongs, carries limitations when, as is the case with pure states, this distance is intrinsically intended for elements of a subset with a structure in which the straight line is not contained, and we mean here the spherical surface of the Bloch sphere to which the pure states are restricted.  
So we must ask ourselves at this point of our analysis which metric is the proper one to be used on pure quantum states. 

\begin{figure}[h!]
\begin{center}
\includegraphics[width=7cm]{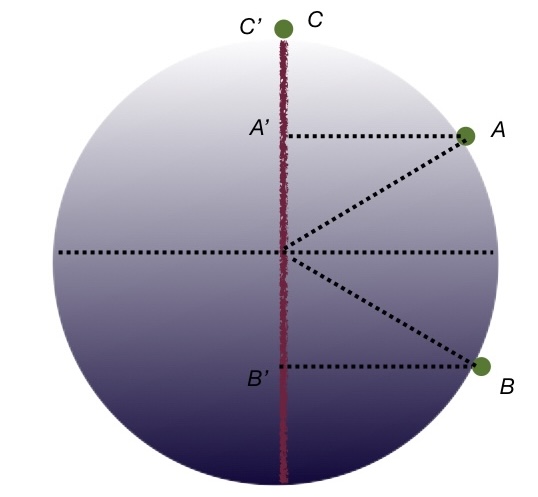}
\end{center}
\caption{We consider a situation where there are two names for colors, which we call {\it Light} and {\it Dark}, and wish to show that the quantum measurement model, which in this case represents a `qubit', incorporates the mechanism of categorical perception. For this, we consider three pure states $|\pi/3, \phi \rangle$, 
$|2\pi/3, \phi \rangle$ and $|0, \phi \rangle$, respectively, in the Bloch representation located in points $A$, $B$ and $C$, which represent stimuli associated with a quantum measurement on this qubit. With each of the three pure states corresponds a density state in which the qubit is located after the measurement, localized in the Bloch representation in respectively points $A'$, $B'$ and $C'$, and described by respectively density matrices (\ref{densitymatrixpi/3}), (\ref{densitymatrix2pi/3}) and (\ref{densitymatrix0}). The pure states in $A$ and $B$ belong to two different colors, {\it Light} and {\it Dark}, and lie at a distance 1/3 from each other. The density states corresponding to them, located in points $A'$ and $B'$, are at a distance 1/2 from each other. We see here the dilation mechanism of categorical reception at work, for percepts belonging to different categories, {\it Light} and {\it Dark}. The pure states in $C$ and $A$ belong to the same color, {\it Light}, and also lie at a distance 1/3 from each other. The density states corresponding to $C$ and $A$, located in points $C'$ and $A'$, are at a distance 1/4 from each other. We see here the contraction mechanism of categorical reception at work, for percepts belonging to the same category.}
\label{LightDarklabel}
\end{figure}

A natural notion of distance between pure states should only account for the pure states that are in-between the points on the Bloch sphere representing these states, the length of the circular arc connecting these points would be such a quantity. More precisely, if $\theta$ and $\alpha$ are the polar angles of the pure states $|\theta, \phi \rangle$ and $|\alpha, \phi \rangle$ (see Figure \ref{LightDarklabel}), the distance to be used to measure how much they are separated from each other, normalized to 1, would then be  
\begin{eqnarray}
    d_{pure}(|\theta, \phi \rangle, |\alpha, \phi \rangle) = {1 \over \pi} |\theta - \alpha|
\end{eqnarray} 
Note that it is sufficient to define the distance for two pure states with equal azimuthal angle $\phi$. Indeed, we can always choose a north pole and a south pole for any two pure states such that both states lie in the plane with equal azimuthal angle, so that then the arc of the circle through both points is determined by the difference of the polar angles of both states.
This distance also results from the angle between pure states calculated from their inner product of the Hilbert space, as was identified already in the early twentieth century by two mathematicians Guido Fubini and Eduard Study to be the natural metric of the projective space of pure states, and it is now called the Fubini Study metric \citep{fubini1904,study1905}. 
For pure states $\psi_1$ and $\psi_2$ the Fubini Study distance is given by 
\begin{eqnarray}
    \gamma(\psi_1,\psi_2) = \arccos |\langle \psi_1 | \psi_2\rangle|
\end{eqnarray}
and normalized it coincides with the arc distance we introduced as natural distance on the pure states in the Bloch sphere.

So, we will use two different notions of distance, one that considers the circular arc between two vector states, at the surface of the Bloch sphere, and the other one which considers the Euclidean distance between density states, inside the Bloch sphere. The normalized to 1 distance between two such decohered density states 
\begin{eqnarray}
\begin{pmatrix}
\cos^2{\theta \over 2} & 0 \\
0 & \sin^2{\theta \over 2}
\end{pmatrix} \quad {\rm and} \quad \begin{pmatrix}
\cos^2{\alpha \over 2} & 0 \\
0 & \sin^2{\alpha \over 2}
\end{pmatrix}
\end{eqnarray}
will be given by (see Figure \ref{LightDarklabel}) 
\begin{eqnarray}
  d_{density}(\theta,\alpha) = {1 \over 2} | \cos\theta - \cos\alpha |
\end{eqnarray}
Equipped with these two distances, let us now analyze how the phenomenon of categorical perception is naturally expressed in a quantum measurement process. As we mentioned, the points of the sphere's diameter where the elastic is stretched are where the percepts lie, for the measurement in question, and the model is Kolmogorovian in that region, the probabilities being an expression of the lack of knowledge of the one who perceives. This is realized in the extended Bloch model by the unpredictable point $\lambda$ where the elastic breaks. Also, the connection between the reality of the considered entity (the pure states describing the stimuli) and the elements of the contextual reality (the on-elastic density states describing the percepts for the given measurement) is illustrated by the deterministic orthogonal fall of the point particle representative of the stimulus onto the elastic, transforming it into a percept, thus reaching a stage where an outcome (an answer) is actualized and we are back to a pure state.

To see how a quantum measurement brings about the warping effect of categorical perception, let us consider, to fix ideas, a situation where only two colors exist, {\it Light} and {\it Dark}, so we are precisely in a situation that can be described in a three-dimensional Bloch sphere. Note that Eleanor Rosch formulated the rationale for the prototype theory for concepts while teaching colors to a primitive community in Papua New Guinea, whose language, called Berinomo, had just two names for colors \citep{rosch1973}. 

Let us locate the first color, {\it Light}, at the North Pole of the Bloch sphere, and the second color, {\it Dark}, at its South Pole. At the equator, the transition from {\it Light} to {\it Dark} will then occur (see Figure \ref{LightDarklabel}). 
Let us then introduce three different pure states, the first one located in point $A$ (see Figure \ref{LightDarklabel}) represented by the vector $|\pi/3, \phi \rangle$, hence with polar angle $\theta = \pi/3$, the second one located in point $B$  (see Figure \ref{LightDarklabel}) represented by the vector $|2\pi/3, \phi \rangle$, hence with polar angle $\theta = 2\pi/3$, and the third one located in the North Pole  (see Figure \ref{LightDarklabel}), represented by the vector $|0, \phi \rangle$, (hence, this is the eigenstate describing {\it Light}, with a polar angle $0$), assuming for simplicity that they all lie on a same plane, hence have the same asimuthal angle $\phi$. When a {\it Light}-{\it Dark} color-measurement is performed, the pre-measurement pure states deterministically transform into the fully decohered pre-collapse density states, obtained by plunging the associated point particles into the sphere, orthogonally with respect to the line subtended by the {\it Light} and {\it Dark} outcome states, which is the region of the percepts, i.e., of the `contextual reality' relative to this specific color-measurement. This results for the pure states located in $A$, $B$ and $C$, respectively, in density states located in points $A'$, $B'$ and $C'$, and given, respectively, by the density matrices 
\begin{eqnarray} \label{densitymatrixpi/3}
&&   \begin{pmatrix}
1/4 & 0 \\
0 & 3/4 
\end{pmatrix}, \\ \label{densitymatrix2pi/3}
&&\begin{pmatrix}
3/4 & 0 \\
0 & 1/4
\end{pmatrix} \\ \label{densitymatrix0}
&& {\rm and} \quad 
\begin{pmatrix}
1 & 0 \\
0 & 0
\end{pmatrix}
\end{eqnarray} 
If we consider the pure states $|{\pi \over 3}, \phi \rangle$ and $|{2\pi \over 3}, \phi \rangle$, located respectively in points $A$ and $B$, they are in the {\it Light} and {\it Dark} hemispheres of the Bloch sphere, hence they are two different colors, and we can now easily see that their transformation to the corresponding decohered density states, located respectively in points $A'$ and $B'$ (see Figure \ref{LightDarklabel}), and described respectively by the density matrices (\ref{densitymatrixpi/3}) and (\ref{densitymatrix2pi/3}) 
exhibits a warping that is a dilation. 
Indeed, their distance is $1/3$, that is, one third of the maximal distance between two stimuli, while the associated percepts, corresponding to the decohered quantum states, have a distance of $1/2$, i.e., one half of the maximal distance between two percepts. On the other hand, if we consider the pure states $|0, \phi \rangle$ and $|{\pi \over 3}, \phi \rangle$, located respectively in points $C$ and $A$,  belonging to the same color, namely the color {\it Light}, the opposite warping occurs. Indeed, on the stimuli side, the distance is again $1/3$, whereas on the percepts side, the corresponding density states being located in points $C'$ and $A'$, respectiveky, and described by the density matrices (\ref{densitymatrix0}) and (\ref{densitymatrixpi/3}), we now have distance $1/4$. This means that a warping takes place, which is now a contraction. This shows how the phenomenon of categorical perception is built into the quantum measurement.
 
\section{Conclusion}
As we illustrated in \citep{aertsaertsarguelles2022}, there are good reasons to believe that the categorical perception mechanism underlies the emergence of concepts for which a prototype exists. Indeed, the mechanism causes compactifications and dilations as a consequence of the kind of warping it causes, and the compactifications lead to the creation of abstractions of the compactifications. These abstractions then become concepts for which the instances can be related in a graded manner with a prototype. In this paper, we have shown that the structure of the quantum measurement process intrinsically contains the categorical perception mechanism. Quantum cognition consists in applying various mathematical structural elements of quantum mechanics to model situations in human cognition. This has already been successfully performed with the quantum structures of `superposition', `entanglement', `non-commutivity' and `indistinguishability'. We refer to \citet{huangetal2025} for an up-to-date overview of the quantum cognition research domain and how by applying the just mentioned quantum structures in models of a multitude of situations from human cognition it systematically provides empirically more adequate descriptions than other cognitive theories, such as, for example, those based on Bayesian probabilities. By demonstrating in this present article that the quantum measurement process intrinsically contains the mechanism of categorical perception, we add to the listed elements of quantum structures a new element, that of `quantum measurement'. Indeed, this element of quantum structure most likely contributes to the success of quantum cognition as a cognitive theory by allowing the basic mechanism of concept formation in human cognition, which is present in literally just about every situation of cognition, to take shape in the proposed quantum cognition models.


\end{document}